\documentstyle[eqsecnum,aps]{revtex}
\begin{document}
\draft
\title{Exact solutions of the Lawrence-Doniach model for layered superconductors}
\author{Sergey V. Kuplevakhsky}
\address{Department of Physics, Kharkov National University,\\
61077 Kharkov, Ukraine\\
and Institute of Electrical Engineering,\\
SAS, 842 39 Bratislava, Slovak Republic}
\date{today}
\maketitle

\begin{abstract}
We solve the problem of exact minimization of the Lawrence-Doniach (LD)
free-energy functional in parallel magnetic fields. We consider both the
infinite in the layering direction case (the infinite LD model) and the
finite one (the finite LD model). We prove that, contrary to a prevailing
view, the infinite LD model does not admit solutions in the form of isolated
Josephson vortices. For the infinite LD model, we derive a closed,
self-consistent system of mean-field equations involving only two variables.
Exact solutions to these equations prove simultaneous penetration of
Josephson vortices into all the barriers, accompanied by oscillations and
jumps of the magnetization, and yield a completely new expression for the
lower critical field. Moreover, the obtained equations allow us to make
self-consistent refinements on such well-known results as the Meissner
state, Fraunhofer oscillations of the critical Josephson current, the upper
critical field, and the vortex solution of Theodorakis [S. Theodorakis,
Phys. Rev. B {\bf 42}, 10172 (1990)]. Our consideration of the finite LD
model illuminates the role of the boundary effect. In contrast to the
infinite case, an explicit analytical solution to the Maxwell equations of
the finite case does not preclude the existence of localized Josephson
vortex configurations. By the use of this solution, we obtain a
self-consistent description of the Meissner state. Finally, we discuss some
theoretical and experimental implications.
\end{abstract}


\section{Introduction}

We obtain exact analytical solutions to the phenomenological
Lawrence-Doniach \cite{LD} (LD) model for layered superconductors in
external parallel magnetic fields. We consider both the infinite in the
layering direction case (the infinite LD model) and the finite one (the
finite LD model). This paper should be considered as a logical continuation
of our previous study of layered superconductors on the basis of a
microscopic approach \cite{K98}.

At present, the LD model is widely used for the description of low-$T_c$
layered superconductors and superlattices as well as high-$T_c$
superconductors exhibiting the intrinsic Josephson effect \cite{T96,KM94}.
Surprisingly, despite a large number of theoretical publications on this
subject, it has not been realized yet that the problem of the parallel
magnetic field is exactly solvable. Up to now, actual analytical solutions
with different degrees of accuracy have been obtained only for relatively
simple particular cases of the infinite LD model: the Meissner state \cite
{BF92}, Fraunhofer oscillations of the critical Josephson current \cite
{BCG92}, the upper critical field $H_{c2}(T)$ \cite{KLB75,T96}, and the
vortex state \cite{Th90} in the intermediate field regime.

Unfortunately, the calculations of the lower critical field $H_{c1}$ \cite
{B73,CCH91}, based on the assumption of isolated Josephson vortex
penetration, raise questions. In these calculations, one employs an
anisotropic continuum approximation outside the so-called ''Josephson vortex
core region'' \cite{CC90}, completely neglecting the intrinsic discreteness
of the LD model. As has been recently shown by Farid \cite{F98}, a set of
equations thus obtained has no physical solution. Furthermore, the
calculations of a triangular Josephson vortex lattice \cite{BC91}, also
based on the assumption of the existence of isolated Josephson vortices, are
at odds with the exact vortex solution of Theodorakis \cite{Th90}, valid in
the same field range and exhibiting full homogeneity in the layering
direction.

As has been shown within the framework of the microscopic theory \cite{K98},
the resolution of these contradictions lies in the analysis of singular
mathematical structure of free-energy functionals of layered
superconductors. In particular, the system of the Maxwell equations in
layered superconductors contains a constraint relation that physically
constitutes the conservation law for the total intralayer current. According
to this constraint relation, the phases of the superconducting order
parameter (the pair potential) at different layers turn out to be mutually
dependent. The minimization of the free energy with respect to the phases
must necessarily take into account this fact. The neglect of mutual
dependence of the phases leads to an incomplete set of mean-field equations.
In the present paper, we elucidate this mathematical issue in full detail.

Section II of the paper is devoted to the infinite in the layering direction
LD model. In subsection II.A, we concentrate on exact minimization of the LD
free-energy functional. Using general field-theoretical arguments, we prove
that the Maxwell equations of the LD model contain an infinite number of
unphysical degrees of freedom that cannot be eliminated by imposing a gauge
condition. We achieve the elimination of these redundant degrees of freedom
by minimizing the free energy with respect to the phases, taking account of
the conservation law for the total intralayer current. In this way, we
obtain a complete set of necessary and sufficient conditions of an
unconditional minimum of the LD functional. These conditions constitute a
remarkably simple, closed, self-consistent system of mean-field equations
involving only two variables: the reduced modulus of the pair potential (the
same for all the superconducting layers) and the phase difference (the same
for all the barriers). In addition, we prove that inhomogeneous in the
layering direction field configurations do not correspond to any stationary
points of the free energy. As a result, contrary to the prevailing view \cite
{B73,CC90,CCH91,BC91}, the infinite LD model does not admit any solutions in
the form of isolated Josephson vortices.

In subsection II.B, we proceed to exact solutions of the mean-field
equations of the infinite LD model describing major physical effects. We
arrive at a new scenario of the flux penetration at $H_{c1}$: We show that
Josephson vortices penetrate all the barriers simultaneously and coherently,
forming homogeneous field distribution in the layering direction (a ''vortex
plane''). The corresponding lower critical field is $H_{c1}=2\left( \pi
ep\lambda _j\right) ^{-1}$, where $p$ is the layering period, $\lambda
_J=\left( 8\pi ej_0p\right) ^{-1/2}$ is the Josephson penetration depth,
with $j_0$ being the critical density of the Josephson current. We show that
the magnetization exhibits oscillations and jumps due to successive vortex
plane penetration. We also obtain all well-known limiting cases [the
Meissner state, Fraunhofer oscillations of the critical Josephson current,
the upper critical field $H_{c2}(T)$, and the vortex solution of
Theodorakis] with self-consistent refinements. All these results stand in
complete agreement with our previous microscopic consideration \cite{K98}.

In section III, we consider the finite, both in the layering direction and
along the layers, LD model. We show that the emergence of additional
boundary conditions in this case completely eliminates unphysical degrees of
freedom of the Maxwell equations and makes minimization with respect to the
phases impossible. An explicit solution to the Maxwell equations obtained in
this section, in contrast to the infinite case, does not preclude the
existence of localized Josephson vortex configurations. As regards the
physical effects, we derive exact analytical expressions for the order
parameter, the currents and the local magnetic field describing the Meissner
state.

In section IV, we present a brief summary of the obtained results and
discuss some theoretical and experimental implications. In appendix A, we
obtain an explicit solution to the Maxwell equations in the infinite case.
We also consider a variation of this solution induced by variations of the
phases. In Appendix B, we discuss relationship to the microscopic theory 
\cite{K98}. This discussion casts light on the actual domain of validity of
the LD model.

\section{The infinite LD model}

In this section, we consider an infinite in the layering direction LD model.
One of the dimensions of the system along the layers is taken to be finite,
although it can be made arbitrarily large.

We begin by reminding basic features of the LD model \cite{LD,KLB75}. In
this model, the temperature $T$ is assumed to be close to the ''intrinsic''
critical temperature $T_{c0}$ of individual layers:

\begin{equation}  \label{1}
\tau \equiv \frac{T_{c0}-T}{T_{c0}}\ll 1.
\end{equation}
The superconducting (S) layers are assumed to have negligible thickness
compared to the ''intrinsic'' coherence length $\zeta (T)\propto \tau
^{-1/2} $, the penetration depth $\lambda (T)\propto \tau ^{-1/2}$, and the
layering period $p$. Taking the layering axis to be $x$, choosing the
direction of the external magnetic field ${\bf H}$ to be $z$ [${\bf H}%
=(0,0,H)$], assuming homogeneity along this axis and setting $\hbar =c=1$,
we can write the LD free-energy functional as

\[
\Omega _{LD}\left[ f_n,\phi _n,\frac{d\phi _n}{dy},A_x,A_y;H\right] =\frac{%
pH_c^2(T)}{4\pi }W_z\int\limits_{L_{y1}}^{L_{y1}}dy\sum_{n=-\infty
}^{+\infty }\left[ -f_n^2(y)+\frac 12f_n^4(y)\right. 
\]

\[
+\zeta ^2(T)\left[ \frac{df_n(y)}{dy}\right] ^2+\zeta ^2(T)\left[ \frac{%
d\phi _n(y)}{dy}-2eA_y(np,y)\right] ^2f_n^2(y) 
\]

\[
\left. +\frac{r(T)}2\left[ f_{n-1}^2(y)+f_n^2(y)-2f_n(y)f_{n-1}(y)\cos \Phi
_{n,n-1}(y)\right] \right] 
\]

\begin{equation}  \label{2}
\left. +\frac{4e^2\zeta ^2(T)\lambda ^2(T)}p\int\limits_{(n-1)p}^{np}dx\left[
\frac{\partial A_y(x,y)}{\partial x}-\frac{\partial A_x(x,y)}{\partial y}-H%
\right] ^2\right] ,
\end{equation}

\[
\Phi _{n,n-1}(y)=\phi _{n,n-1}(y)-2e\int\limits_{(n-1)p}^{np}dxA_x(x,y), 
\]
\[
\phi _{n,n-1}(y)=\phi _n(y)-\phi _{n-1}(y). 
\]
Here ${\bf A}=(A_x,A_y,0)$ is the vector potential, continuous at the
S-layers: ${\bf A(}np-0,y)={\bf A(}np+0,y)={\bf A(}np,y)$; $W_z$ is the
length of the system in the $z$ direction ($W_z\rightarrow \infty $); $%
f_n(y) $ [$0\leq f_n(y)\leq 1$] and $\phi _n(y)$ are, respectively, the
reduced modulus and the phase of the pair potential $\Delta _n(y)$ in the $n$%
th superconducting layer: 
\begin{equation}  \label{3}
\Delta _n(y)=\Delta (T)f_n(y)\exp \phi _n(y),
\end{equation}
with $\Delta (T)$ being the ''intrinsic'' gap [$\Delta (T)\propto \tau
^{1/2} $]; $H_c(T)$ is the thermodynamic critical field; $r(T)=2\alpha
_{ph}\tau ^{-1}$ is a dimensionless phenomenological parameter of the
Josephson interlayer coupling ($0<\alpha _{ph}\ll 1$). The local magnetic
field ${\bf h}=(0,0,h)$ obeys the relation 
\begin{equation}  \label{4}
h(x,y)=\frac{\partial A_y(x,y)}{\partial x}-\frac{\partial A_x(x,y) }{%
\partial y}.
\end{equation}

\subsection{Exact minimization of the LD functional}

Our task now is to establish a closed, complete, self-consistent system of
mean-field equations for the pair potential $\Delta _n$ and the local
magnetic field $h$, which is mathematically equivalent to the minimization
of (\ref{2}) with respect to $f_n$, $\phi _n$, and ${\bf A}$. This problem
should be approached with a great deal of caution because of singular
mathematical structure of the functional (\ref{2}), resulting from gauge
invariance combined with discreteness. Thus, one must take account of the
fact that variations with respect to $\phi _n$ and ${\bf A}$ are not
independent. Moreover, variations with respect to $\phi _n$ at different
layers in themselves turn out to be mutually dependent. Unfortunately, these
crucial points have not been realized in previous literature. To clarify
them, we consider partial variational derivatives with respect to $\phi _n$,
and $A_{x\text{, }}A_y$, formally obtained under the assumption of the
independence of these variables.

As the functional (\ref{2}) is invariant under the gauge transformation 
\[
\phi _n(y)\rightarrow \phi _n(y)+2e\eta (np,y),\text{ }A_i(x,y)\rightarrow
A_i(x,y)+\partial _i\eta (x,y),\text{ }i=x,y, 
\]
where $\eta (x,y)$ is an arbitrary smooth function of $x$, $y$ in the whole
region $\left( -\infty <x<+\infty \right) \times \left[ L_{y1}\leq y\leq
L_{y2}\right] $, partial variational derivatives with respect to $\phi _n$,
and $A_{x\text{, }}A_y$ are related by the fundamental identities 
\begin{equation}  \label{6}
2e\frac{\delta \Omega _{LD}}{\delta \phi _n(y)}\equiv \frac \partial {%
\partial y}\frac{\delta \Omega _{LD}}{\delta A_y(np,y)}+\frac{\delta \Omega
_{LD}}{\delta A_x(np+0,y)}-\frac{\delta \Omega _{LD}}{\delta A_x(np-0,y)}.
\end{equation}
Being a consequence of Noether's second theorem, such identities are typical
of any gauge theory \cite{KP80}. They imply that the number of independent
Euler-Lagrange equations is less than the number of variables. In other
words, the system of Euler-Lagrange equations contains unphysical degrees of
freedom whose number is equal to the number of Noether's identities. Unusual
is, however, an infinite number of identities (\ref{6}). Indeed, in
continuum gauge theories [such as, e.g., the Ginzburg-Landau (GL) theory of
superconductivity] the number of Noether's identities is equal to the number
of independent parameters of the relevant gauge group. [In the case of
superconductivity, we are dealing with the electromagnetic one-parameter
group $U(1).$] Thus, by imposing gauge conditions in continuum gauge
theories, one completely eliminates all unphysical degrees of freedom. By
contrast, in the discrete LD theory a single available gauge condition
cannot eliminate an infinite number of unphysical degrees of freedom
resulting from (\ref{6}). The resolution of the problem of the remaining
''infinity minus one'' unphysical degrees of freedom lies in implicit mutual
dependence of the variations with respect to the phases $\phi _n$ at
different S-layers. Below, we demonstrate this dependence explicitly. [See
relation (\ref{16}).]

To finish with the discussion of (\ref{6}), we point out that these same
identities hold also for the LD model with decoupled S-layers [when $%
r(T)\equiv 0$]. However, now the number of unphysical degrees of freedom is
equal to the number of physically independent systems [one identity (\ref{6}%
) per independent S-layer]. A single gauge condition completely eliminates
the arbitrariness of the Euler-Lagrange equations in this case.

We start by minimizing with respect to ${\bf A}$. Varying (\ref{2}) with
respect to $A_x$, $A_y$ in the regions $(n-1)p<x<np$ under the condition $%
\delta A_x(x,L_{y1})=\delta A_x(x,L_{y2})=0$ yields the Maxwell equations 
\begin{equation}  \label{9}
\frac{\partial h(x,y)}{\partial y}=4\pi j_{n,n-1}(y)\equiv 4\pi
j_0f_n(y)f_{n-1}(y)\sin \Phi _{n,n-1}(y),
\end{equation}
\begin{equation}  \label{10}
\frac{\partial h(x,y)}{\partial x}=0,
\end{equation}
where $j_{n,n-1}(y)$ is the density of the Josephson current between the $%
(n-1)$th and the $n$th layers, $j_0=r(T)p/16\pi e\zeta ^2(T)\lambda ^2(T)$.
Minimization with respect to $A_y(np,y)$ leads to boundary conditions at the
S-layers 
\begin{equation}  \label{11}
h(np-0,y)-h(np+0,y)=\frac{pf_n^2(y)}{2e\lambda ^2(T)}\left[ \frac{d\phi _n(y)%
}{dy}-2eA_y(np,y)\right] .
\end{equation}

Equations (\ref{9})-(\ref{11}) should be complemented by boundary conditions
at the outer interfaces $y=L_{y1},L_{y2}$. As we do not consider here
externally applied currents in the $y$ direction, the first set of boundary
conditions follows from the requirement that the intralayer currents vanish
at $y=L_{y1},L_{y2}$: 
\begin{equation}  \label{12}
\left[ \frac{d\phi _n(y)}{dy}-2eA_y(np,y)\right] _{y=L_{y1},L_{y2}}=0.
\end{equation}
Applied to Eqs. (\ref{11}), these boundary conditions show that the local
magnetic field at the outer interfaces is independent of the coordinate $x$: 
$h(x,L_{y1})=h(L_{y1})$, $h(x,L_{y2})=h(L_{y2})$. Boundary conditions
imposed on $h$ should be compatible with Ampere's law $h(L_{y2})-h(L_{y1})=4%
\pi I$ obtained by integration of Eqs. (\ref{9}) over $y$, where 
\begin{equation}  \label{13}
I\equiv
\int\limits_{L_{y1}}^{L_{y2}}dyj_{n+1,n}(y)=\int%
\limits_{L_{y1}}^{L_{y2}}dyj_{n,n-1}(y)
\end{equation}
is the total Josephson current.

Differentiating (\ref{11}) with respect to $y$ and employing (\ref{9}), we
arrive at the current-continuity laws for the S-layers: 
\[
\frac \partial {\partial y}\left[ f_n^2(y)\left[ \frac{d\phi _n(y)}{dy}%
-2eA_y(np,y)\right] \right] 
\]
\begin{equation}  \label{15}
=\frac{r(T)}{2\zeta ^2(T)}f_n(y)\left[ f_{n-1}(y)\sin \Phi
_{n,n-1}(y)-f_{n+1}(y)\sin \Phi _{n+1,n}(y)\right] .
\end{equation}
These relations may be interpreted as ''the Euler-Lagrange equations for the
phases'' in the sense that they can be formally obtained by taking partial
variational derivatives with respect to $\phi _n$ under conditions (\ref{12}%
). However, actual minimization of (\ref{2}) with respect to the phases must
take account of mutual dependence of $\delta \phi _n(y)$ at different
layers, as shown in what follows. [The fact that relations (\ref{15}) follow
directly from the Maxwell equations (\ref{9}), (\ref{11}) is a consequence
of (\ref{6}). Surprisingly, this trivial functional dependence of the
current-continuity laws for the S-layers on the Maxwell equations has not
been pointed out in the previous literature.\cite{BLK92}]

Adding Eqs. (\ref{15}), integrating and using boundary conditions (\ref{12}%
), we get the conservation law for the total intralayer current: 
\begin{equation}  \label{16}
\sum_{n=-\infty }^{+\infty }f_n^2(y)\left[ \frac{d\phi _n(y)}{dy}-2eA_y(np,y)%
\right] =0.
\end{equation}
This key relation of our consideration has mathematical form of a constraint 
\cite{L62} on the derivatives of the phases and the $y$ components of the
vector potential at different S-layers. Unfortunately, the existence of the
constraint relation (\ref{16}) in the system of the Maxwell equations (\ref
{9})-(\ref{11}) has not been noticed in previous publications, hence
difficulties in establishing a complete set of necessary and sufficient
conditions of an unconditional minimum of (\ref{2}). We want to emphasize
that the fundamental constraint relation (\ref{16}) and its corollaries
below [relations (\ref{19}), (\ref{18})] should not be confused with
auxiliary constraint relations imposed on independent variables in the
standard variational problem of a conditional minimum \cite{L62}. All
constraints of the LD model appear as a result of singular structure of the
functional (\ref{2}) itself. (See Refs. \cite{D64,GT86} for a thorough
discussion of singular field theories of this type.)

According to main principles of the calculus of variations \cite{L62}, to
minimize (\ref{2}) with respect to $\phi _{n}$, we must first eliminate the
constraint (\ref{16}). Assuming that $f_{m}(y)>0$, where $m$ is an arbitrary
layer index, we rewrite (\ref{16}) as 
\begin{equation}
2eA_{y}(mp,y)=\frac{d\phi _{m}(y)}{dy}+\frac{1}{f_{m}^{2}(y)}\sum_{n\neq
m}f_{n}^{2}(y)\left[ \frac{d\phi _{n}(y)}{dy}-2eA_{y}(np,y)\right] .
\label{17}
\end{equation}
Equation (\ref{17}) expresses $A_{y}(mp,y)$ as a function of all $\frac{%
d\phi _{n}(y)}{dy}$. It should be substituted into (\ref{2}). Now all $%
\delta \phi _{n}(y)$ can be considered as independent. Carrying out the
variation under the conditions (\ref{12}), we obtain 
\begin{equation}
f_{m-1}(y)\sin \Phi _{m,m-1}(y)-f_{m+1}(y)\sin \Phi _{m+1,m}(y)=0,
\label{19}
\end{equation}
\[
\frac{\partial }{\partial y}\left[ f_{n}^{2}(y)\left[ \frac{d\phi _{n}(y)}{dy%
}-2eA_{y}(np,y)\right] \right] -\frac{\partial }{\partial y}\left[
f_{n}^{2}(y)\left[ \frac{d\phi _{m}(y)}{dy}-2eA_{y}(mp,y)\right] \right] 
\]
\[
=\frac{r(T)}{2\zeta ^{2}(T)}f_{n}(y)\left[ f_{n-1}(y)\sin \Phi
_{n,n-1}(y)-f_{n+1}(y)\sin \Phi _{n+1,n}(y)\right] ,\quad n\not=m.
\]
Comparing these equations with (\ref{15}) and integrating with boundary
conditions (\ref{12}) for $n=m$ yields 
\begin{equation}
\frac{d\phi _{m}(y)}{dy}-2eA_{y}(mp,y)=0.  \label{18}
\end{equation}
Since $m$ is an arbitrary layer index, relations (\ref{19}), (\ref{18}) hold
for all $n=m=0,\pm 1,\pm 2,\ldots $ Note that only one of the two sets of
relations (\ref{19}), (\ref{18}) is independent. For example, relations (\ref
{19}) can be obtained by inserting (\ref{18}) into (\ref{15}) and vice
versa. In turn, the number of independent relations (\ref{18}) is exactly
equal to ''infinity minus one'', because they obey the constraint (\ref{16}%
). As expected, the correct minimization of (\ref{2}) with respect to the
phases completely resolves the problem of unphysical degrees of freedom
contained in Eqs. (\ref{9})-(\ref{11}). Physically, relations (\ref{18}),
which appear already in the case of decoupled layers, minimize the kinetic
energy of the intralayer currents and, by (\ref{11}), assure the continuity
of the local magnetic field at the S-layers. [According to (\ref{10}), $h$
does not depend on $x$ in the barrier regions. Thus, $h(x,y)=h(y)$ in the
whole region $\left( -\infty <x<+\infty \right) \times \left[ L_{y1}\leq
y\leq L_{y2}\right] $.] Relations (\ref{19}) constitute stationarity
conditions for the Josephson term in (\ref{2}) and assure the continuity of
the Josephson current at the S-layers as required by (\ref{13}).

The above results, in fact, prove that inhomogeneous in the layering
direction field configurations [i.e., those that do not satisfy (\ref{19}), (%
\ref{18})] do not correspond to any stationary points of the functional (\ref
{2}). Consider the variation of the solution of (\ref{9})-(\ref{12}) for $%
A_y $ in the gauge $A_x=0$ on an interval $(m-1)p<x\leq mp$, induced by a
variation of the phase at the $n$th layer. According to (\ref{a4}), we have $%
\delta A_y(x,y)=\frac 1{2e}\frac{f_n^2(y)}{f_m^2(y)}\frac{d\delta \phi _n(y) 
}{dy}$. Such a variation does not affect the energy of the magnetic field in
(\ref{2}). If $n=m$, the variation of the kinetic energy of the intralayer
currents vanishes, but the first-order variation of the Josephson term is
nonzero. If $n\not =m$, the variation of the Josephson term vanishes, but
now the first-order variation of the kinetic energy of the intralayer
currents is nonzero. These first-order variations of (\ref{2}) vanish if and
only if the conditions (\ref{19}), (\ref{18}) are fulfilled (i.e., for
homogeneous field configurations). Unfortunately, this general mathematical
consideration unambiguously precludes the existence of isolated Josephson
vortices \cite{B73,CCH91,CC90,BC91} in the infinite LD model. It also
explains the results of Farid \cite{F98}, who has pointed out
inconsistencies in a mathematical description of such hypothetical entities.

It is instructive to look at the incompleteness of the system (\ref{9})-(\ref
{11}) from a slightly different mathematical point of view. In the gauge $%
A_x=0$, this system reduces to an infinite set of integrodifferential
equations (\ref{a2}) for the phase differences $\phi _{n,n-1}$ (for fixed $%
f_n$). There are no theorems of existence and uniqueness of a solution to an
infinite set of such equations. By contrast, for a finite set, describing a
finite in the layering direction layered superconductor, the existence and
uniqueness of a solution can be proved by standard methods of functional
analysis. The description of a finite layered superconductor implies the
specification of boundary conditions on {\bf A} at the ''top'' and
''bottom'' S-layers, whereas the infinite LD model considered here does not
impose any boundary conditions on {\bf A} at{\bf \ }$x\rightarrow \pm \infty 
$. Thus, the arbitrariness contained in Eqs. (\ref{9})-(\ref{11}) is an
intrinsic mathematical property, necessary to satisfy additional boundary
conditions in the case of the finite LD model. This issue is discussed in
more detail in section III.

Minimization with respect to $f_n$ is straightforward. Under the condition
that $\delta f_n(L_{y1})$, $\delta f_n(L_{y2})$ are arbitrary, we get 
\[
f_n(y)-f_n^3(y)+\zeta ^2(T)\frac{d^2f_n(y)}{dy^2} 
\]
\[
=\frac{r(T)}2\left[ 2f_n(y)-f_{n+1}(y)\cos \Phi _{n+1,n}(y)-f_{n-1}(y)\cos
\Phi _{n,n-1}(y)\right] 
\]
\begin{equation}  \label{31}
+\zeta ^2(T)\left[ \frac{d\phi _n(y)}{dy}-2eA_y(np,y)\right] ^2f_n(y),
\end{equation}
\begin{equation}  \label{32}
\frac{df_n}{dy}\left( L_{y1}\right) =\frac{df_n}{dy}\left( L_{y2}\right) =0.
\end{equation}

Equations (\ref{9})-(\ref{11}), (\ref{31}) and (\ref{19}) [or, equivalently,
(\ref{18})] (with $m\rightarrow n$), together with boundary conditions (\ref
{12}), (\ref{11}) and boundary conditions for $h(y)$, form a closed,
complete set of necessary and sufficient conditions of all the stationary
points of the functional (\ref{2}). For example, the well-known maximum $%
\Omega _{LD}=0$ for $H=I=0$ (the normal state) trivially satisfies these
conditions with $f_n=0$. The absolute minimum $\Omega _{LD}=-\frac{H_c^2(T)V 
}{8\pi }$ ($V$ is the volume of the system) for $H=I=0$ also satisfies these
conditions with $\Phi _{n+1,n}=0$ and $f_n=1$. Complemented by the
requirement that the Josephson term be a minimum, these conditions become
necessary and sufficient conditions of all the minima of (\ref{2}) for $H%
\not =0$, $I\not =0,$ provided that $\Omega _{LD}<0$. (For $%
L_{y2}-L_{y1}<+\infty $, the Josephson term is bounded and thus has both
minimum and maximum values.)

Indeed, the Josephson term is minimized automatically. The kinetic energy of
the intralayer currents is minimized by (\ref{18}). The energy of the
magnetic field [the last term in (\ref{2})] reaches its minimum value for
given $H$ and $I$ too. This term is non-negative and necessarily has a
minimum determined by the condition that its first-order variation vanish.
(No other stationary points are available.) In the gauge $A_x=0$, the
first-order variation of the magnetic-field energy has the form 
\[
\delta \Omega _{LD}^{mf}\left[ A_y;H\right] =\frac{e^2H_c^2(T)\zeta
^2(T)\lambda ^2(T)W_z}\pi \int\limits_{L_{y1}}^{L_{y2}}dy\sum_{n=-\infty
}^{+\infty }\left\{ -\int\limits_{(n-1)p}^{np}dx\frac{\partial ^2A_y(x,y)}{%
\partial x^2}\delta A_y(x,y)\right. 
\]
\begin{equation}  \label{33}
\left. +\left[ \frac{\partial A_y}{\partial x}\left( np-0,y\right) -\frac{%
\partial A_y}{\partial x}\left( np+0,y\right) \right] \delta
A_y(np,y)\right\} .
\end{equation}
The vanishing of the volume variation in (\ref{33}) (the first term on the
right-hand side) is assured by the Maxwell equations (\ref{10}). The surface
variation [the second term on the right-hand side of (\ref{33})] vanishes by
virtue of (\ref{11}) and (\ref{18}). Consider now the condensation energy in
(\ref{2}) (the sum of the first three phase- and field-independent terms).
This energy reaches its absolute minimum for $f_n=1$, i.e. when the
right-hand side of (\ref{32}) is identically equal to zero. The Josephson
term and the kinetic energy of the intralayer currents induce spatial
dependence and a reduction of $f_n$, which increases the condensation
energy. This influence is minimized under the considered conditions: the
second term on the right-hand side of (\ref{32}) vanishes according (\ref{18}%
) and the first term is minimal when the Josephson energy is a minimum.

Thus, we have proved that the above obtained conditions minimize all the
terms of the functional (\ref{2}): the condensation energy, the Josephson
energy, the kinetic energy of the intralayer currents and the magnetic-field
energy. Any deviation from a solution satisfying these conditions increases
all these terms. As a result, the overall LD free energy increases, as
should be the case for an unconditional minimum \cite{L62}.

Now we proceed to the simplification of Eqs. (\ref{9})-(\ref{11}), (\ref{31}%
) and (\ref{19}) (with $m\rightarrow n$). As the local magnetic field $h$
does not depend on $x$ in the whole region $\left( -\infty <x<+\infty
\right) \times \left[ L_{y1}\leq y\leq L_{y2}\right] $, the quantities $f_n$%
, $\Phi _{n,n-1}$ cannot depend on the layer index: 
\begin{equation}  \label{20}
f_n(y)=f_{n-1}(y)=f(y),\text{ }\Phi _{n+1,n}(y)=\Phi _{n,n-1}(y)=\Phi (y).
\end{equation}
The remaining unphysical degree of freedom of Eqs. (\ref{9})-(\ref{11}), (%
\ref{19}), related to the gauge invariance, is eliminated by fixing the
gauge: 
\begin{equation}  \label{21}
A_x(x,y)=0,\text{ }A_y(x,y)\equiv A(x,y).
\end{equation}
[Note that $\partial $$A/\partial x$ and $\partial ^2$$A/\partial x\partial
y $ are continuous at the S-layers by virtue of (\ref{11}), (\ref{18}), and (%
\ref{9}), (\ref{19}).] The second set of relations (\ref{20}) now yields $%
\phi _n(y)=n\phi (y)+\eta (y)$, where $\phi (y)$ is the coherent phase
difference (the same at all the barriers), and $\eta (y)$ is an arbitrary
function of $y$ that can be set equal to zero without any loss of generality.

From (\ref{10}), employing the continuity conditions for $A,$ $\partial
A/\partial x$ and relations (\ref{18}), we obtain 
\begin{equation}  \label{34}
A(x,y)=\frac 1{2ep}\frac{d\phi (y)}{dy}x.
\end{equation}
Making use of these results$,$ we reduce the functional (\ref{2}) to 
\[
\Omega _{LD}\left[ f,\phi ;H\right] =\frac{H_c^2(T)}{4\pi }%
W_xW_z\int\limits_{L_{y1}}^{L_{y2}}dy\left[ -f^2(y)+\frac 12f^4(y)+\zeta
^2(T)\left[ \frac{df(y)}{dy}\right] ^2\right. 
\]
\begin{equation}  \label{30}
\left. +r(T)\left[ 1-\cos \phi (y)\right] f^2(y)+4e^2\zeta ^2(T)\lambda ^2(T)%
\left[ \frac 1{2ep}\frac{d\phi (y)}{dy}-H\right] ^2\right] ,
\end{equation}
where $W_x=L_{x2}-L_{x1}$. The desired closed, self-consistent set of
mean-field equations for the pair potential $\Delta _n(y)$ and the local
magnetic field $h(y)$ takes the form 
\begin{equation}  \label{22}
\Delta _n(y)=\Delta f(y)\exp \left[ in\phi (y)\right] ,
\end{equation}
\begin{equation}  \label{23}
f(y)+\zeta ^2(T)\frac{d^2f(y)}{dy^2}-f^3(y)-r(T)\left[ 1-\cos \phi (y)\right]
f(y)=0,
\end{equation}
\begin{equation}  \label{24}
\frac{df}{dy}(L_{y1})=\frac{df}{dy}(L_{y2})=0,
\end{equation}
\begin{equation}  \label{25}
\frac{d^2\phi (y)}{dy^2}=\frac{f^2(y)}{\lambda _J^2}\sin \phi (y),
\end{equation}
\begin{equation}  \label{26}
\lambda _J=\left( 8\pi ej_0p\right) ^{-1/2},
\end{equation}
\begin{equation}  \label{27}
h(y)=\frac 1{2ep}\frac{d\phi (y)}{dy},
\end{equation}
\begin{equation}  \label{28}
j(y)\equiv j_{n,n-1}(y)\equiv j_0f^2(y)\sin \phi (y)=\frac 1{4\pi } \frac{%
dh(y)}{dy},
\end{equation}
where $h(y)$ should satisfy appropriate boundary conditions at $%
y=L_{y1},L_{y2}$ with $I\equiv \int\limits_{L_{y1}}^{L_{y2}}dyj(y)$ [see Eq.
(\ref{13}) above].

Remarkably, the coherent phase difference $\phi $ (the same for all the
barriers) obeys only one nonlinear second-order differential equation (\ref
{25}) with only one length scale, the Josephson penetration depth $\lambda
_J $ [Eq. (\ref{26})], as in the case of the Ferrell-Prange equation for a
single junction \cite{BP82}. [Mathematically, equation (\ref{25}) is a
solvability condition for the Maxwell equations.] Due to the factor $f^2$,
equation (\ref{25}) is coupled to nonlinear second-order differential
equation (\ref{23}) describing the spatial dependence of the superconducting
order parameter $f$ (the same for all the S-layers). Equations (\ref{24})
constitute boundary conditions for (\ref{23}). The Maxwell equations (\ref
{27}), (\ref{28}), combined together, yield Eq. (\ref{25}), as they should
by virtue of self-consistency.

It is important to note that equations (\ref{22})-(\ref{28}), with an
appropriate microscopic identification of $r(T)$ and $j_0$, can be
considered as a limiting case of the true microscopic equations \cite{K98}.
(See Appendix B for more details.)

Equations (\ref{23})-(\ref{28}), together with (\ref{30}), encompass the
whole physics of the infinite LD model in parallel magnetic fields. They
admit exact analytical solutions for all physical situations of interest.
These solutions are discussed in the next subsection.

\subsection{Major physical effects}

\subsubsection{\bf The Meissner state}

Consider a semi-infinite (in the $y$ direction) LD superconductor with $%
r(T)\ll 1$, $L_{y1}=0$, $L_{y2}\rightarrow +\infty $ in the external fields 
\begin{equation}  \label{1.35}
0\leq H\leq H_s=(ep\lambda _J)^{-1},
\end{equation}
In the Meissner state, $j(y)\rightarrow 0$, $h(y)\rightarrow 0$ for $%
y\rightarrow +\infty $. The requirement that the Josephson term in (\ref{30}%
) be a minimum means that the density of the Josephson energy should vanish
at $y\rightarrow +\infty $. This leads to the boundary conditions 
\begin{equation}  \label{36}
\frac{d\phi }{dy}\left( 0\right) =2epH,\quad \frac{d\phi }{dy}\left( +\infty
\right) =0,\quad \phi (+\infty )=0,\quad f(+\infty )=1.
\end{equation}
The solution of Eqs. (\ref{23}), (\ref{25}), (\ref{27}), (\ref{28}), subject
to (\ref{24}) and (\ref{36}), up to first order in the small parameter $r(T)$
has the form 
\begin{equation}  \label{2.37}
\phi (y)=-4\arctan \frac{H\exp \left[ -\frac y{\lambda _J}\right] }{H_s+%
\sqrt{H_s^2-H^2}},
\end{equation}
\begin{equation}  \label{3.38}
h(y)=\frac{2HH_s\left[ H_s+\sqrt{H_s^2-H^2}\right] \exp \left[ -\frac y{%
\lambda _J}\right] }{\left[ H_s+\sqrt{H_s^2-H^2}\right] ^2+H^2\exp \left[ -%
\frac{2y}{\lambda _J}\right] },
\end{equation}
\[
j(y)=-\frac{HH_s}{2\pi \lambda _J}\left[ H_s+\sqrt{H_s^2-H^2}\right] 
\]
\begin{equation}  \label{4.39}
\times \frac{\left[ \left[ H_s+\sqrt{H_s^2-H^2}\right] ^2-H^2\exp \left[ -%
\frac{2y}{\lambda _J}\right] \right] \exp \left[ -\frac y{\lambda _J}\right] 
}{\left[ \left[ H_s+\sqrt{H_s^2-H^2}\right] ^2+H^2\exp \left[ -\frac{2y}{%
\lambda _J}\right] \right] ^2},
\end{equation}
\begin{equation}  \label{5.40}
f(y)=1-4r(T)\frac{H^2\left[ H_s+\sqrt{H_s^2-H^2}\right] ^2\exp \left[ -\frac{%
2y}{\lambda _J}\right] }{\left[ \left[ H_s+\sqrt{H_s^2-H^2}\right]
^2+H^2\exp \left[ -\frac{2y}{\lambda _J}\right] \right] ^2}.
\end{equation}
The Meissner solution persists up to the field $H_s=(ep\lambda _J)^{-1}$
that should be regarded as the superheating field of the Meissner state.
This fact was established for the LD model by Buzdin and Feinberg \cite{BF92}%
. A self-consistent solution of the type (\ref{2.37})-(\ref{5.40}) was first
obtained in the framework of the microscopic theory \cite{K98}. In fields $%
H>H_s$, only vortex solutions are possible.

\subsubsection{{\bf The lower critical field $H_{c1\infty }$. Vortex planes}}

Consider now an infinite (in the $y$ direction) LD superconductor with $%
r(T)\ll 1$, $L_{y1}\rightarrow -\infty $, $L_{y2}\rightarrow +\infty $, and $%
j(y)\rightarrow 0$, $h(y)\rightarrow 0$ for $y\rightarrow \pm \infty $. We
are interested in topological solutions of Eqs. (\ref{23}), (\ref{25}), (\ref
{27}), (\ref{28}) for this situation. The requirement that the Josephson
term be a minimum should now be understood as the condition that the density
of the Josephson energy vanish at $y\rightarrow \pm \infty $. Thus, the
appropriate boundary conditions are

\begin{equation}  \label{6.41}
\phi (-\infty )=0,\quad \phi (+\infty )=\pm 2\pi ,\quad \frac{d\phi }{dy}%
(\pm \infty )=0,\quad f\left( \pm \infty \right) =1.
\end{equation}
[Note that aside from $\phi (+\infty )-\phi (-\infty )=\pm 2\pi $ no other
topological boundary conditions are possible. This fact can be proved
analogously to the well-known case of the sine-Gordon model \cite{R82}.]

The desired solutions up to first order in the small parameter $r(T)$ are
given by 
\begin{equation}  \label{7.42}
\phi (y)=\pm 4\arctan \exp \left[ \frac y{\lambda _J}\right] ,
\end{equation}
\begin{equation}  \label{8.43}
h(y)=\pm H_s\cosh {}^{-1}\left[ \frac y{\lambda _J}\right] ,
\end{equation}
\begin{equation}  \label{9.44}
j(y)=\mp 2j_0\cosh {}^{-2}\left[ \frac y{\lambda _J}\right] \sinh {}\left[ 
\frac y{\lambda _J}\right] ,
\end{equation}
\begin{equation}  \label{10.45}
f(y)=1-4r(T)\frac{\exp \left[ -\frac{2\left| y\right| }{\lambda _J}\right] }{%
\left[ 1+\exp \left[ -\frac{2\left| y\right| }{\lambda _J}\right] \right] ^2}%
.
\end{equation}
These solutions explicitly satisfy the usual conditions of the phase and
flux quantization. Indeed, consider a closed rectangular contour $\Gamma $
joining the points $\left( -\frac N2p,-\infty \right) $, $\left( -\frac N2%
p,+\infty \right) ,$ $\left( +\frac N2p,+\infty \right) $ and $\left( +\frac %
N2p,-\infty \right) $. The total change of the phase along this contour for
the ''plus'' sign in (\ref{7.42}) is 
\[
\Delta _\Gamma \phi =\int\limits_{-\infty }^{+\infty }dy\frac{d\phi _{+\frac %
N2}(y)}{dy}+\int\limits_{+\infty }^{-\infty }dy\frac{d\phi _{-\frac N2}(y)}{%
dy}=2\pi N. 
\]
Analogously, the total flux through this contour is 
\[
\Phi _\Gamma =Np\int\limits_{-\infty }^{+\infty }dyh(y)=N\Phi _0, 
\]
where $\Phi _0=\pi /e$ is the flux quantum. Thus, the solution with the
''plus'' sign describes a chain of Josephson vortices positioned in the
plane $y=0$ (one vortex per each barrier). Such a solution was first
obtained in the framework of the microscopic theory\cite{K98} and termed ''a
vortex plane''. The solution with the ''minus'' sign in (\ref{7.42})
describes a chain of Josephson antivortices in the plane $y=0$ (i.e., ''an
antivortex plane'').

By inserting (\ref{7.42}) with the ''plus'' sign and (\ref{10.45}) into (\ref
{30}) and comparing the result with the free energy of the Meissner state,
we derive the lower critical field $H_{c1\infty }$, at which the
vortex-plane solution becomes energetically favorable: 
\begin{equation}  \label{11.46}
H_{c1\infty }=\frac 2\pi H_s=\frac 2\pi \frac{\Phi _0}{\pi p\lambda _J}.
\end{equation}
Note that $h(0)=H_s>H_{c1\infty }$. This means that the penetration of
Josephson vortices at fields $H_{c1\infty }<H<H_s$ can be prevented by a
surface barrier, which should result in hysteretic behavior of magnetization 
\cite{K98}. Finally, we point out that simultaneous Josephson vortex
penetration, envisaged by the vortex-plane solution, and hysteresis in the
magnetization have recently been observed experimentally on artificial
low-temperature superconducting superlattices Nb/Si \cite{Yu98}.

\subsubsection{\bf The vortex state in intermediate fields}

Now we turn to finite-size (in the $y$ direction) LD superconductors with $%
r(T)\ll 1$, $-L_{y1}=L_{y2}\equiv W/2$, in the field range $H_s\ll H\ll
H_{c2\infty }$ ($H_{c2\infty }$ is the upper critical field) and in the
absence of externally applied current ($I=0$). The boundary conditions on $%
\phi $ have the form 
\begin{equation}  \label{47}
\frac 1{2ep}\frac{d\phi }{dy}\left( \pm \frac W2\right) =H.
\end{equation}

Under these conditions, the phase difference up to first order in the small
parameter $H_s^2/H^2$ is 
\begin{equation}  \label{12.48}
\phi (y)=2epHy+\pi N_v(H)-\frac{(-1)^{N_v}}4\frac{H_s^2}{H^2}\left[ \sin
\left( 2epHy\right) -2epHy\cos \left( epWH\right) \right] .
\end{equation}
The constant of integration $\pi N_v(H)$ accounts for the requirement that
the Josephson term in the free energy be a minimum. The ''topological
index'' $N_v$ corresponds to the number of vortex planes and is a singular
function of the applied field $H$: 
\begin{equation}  \label{13.49}
N_v(H)=\left[ \frac{epWH}\pi \right] =\left[ \frac \Phi {\Phi _0}\right] .
\end{equation}
Here $\left[ u\right] $ means the integer part of $u$, and $\Phi =pWH$ is
the flux through one barrier.

By the use of (\ref{12.48}), we derive the following expressions for the
physical quantities up to first order in the small parameters $r(T)$ and $%
H_s^2/H^2$: 
\begin{equation}  \label{15.50}
h(y)=H\left[ 1-\frac{(-1)^{N_v}}4\frac{H_s^2}{H^2}\left[ \cos \left(
2epHy\right) -\cos \left( epWH\right) \right] \right] ,
\end{equation}
\begin{equation}  \label{16.51}
j(y)=(-1)^{N_v}j_0\sin \left( 2epHy\right) ,
\end{equation}
\begin{equation}  \label{17.52}
f(y)=1-\frac{r(T)}2\left[ 1-\frac{(-1)^{N_v}\cos \left( 2epHy\right) }{1+2%
\left[ ep\zeta (T)H\right] ^2}-\frac{\sqrt{2}ep\zeta (T)H\left| \sin \left(
epWH\right) \right| }{1+2\left[ ep\zeta (T)H\right] ^2}\frac{\cosh \frac{%
\sqrt{2}y}{\zeta (T)}}{\sinh \frac W{\sqrt{2}\zeta (T)}}\right] .
\end{equation}

In the limit $W\gg \zeta (T)$, $\left| y\right| \ll W/2$, equation (\ref
{17.52}) becomes 
\begin{equation}  \label{18.53}
f(y)=1-\frac{r(T)}2\left[ 1-\frac{(-1)^{N_v}\cos \left( 2epHy\right) }{1+2%
\left[ ep\zeta (T)H\right] ^2}\right] .
\end{equation}
The vortex solution (\ref{12.48}), (\ref{15.50}), (\ref{18.53}) for $N_v=2m$
($m$ is an integer) was first obtained by Theodorakis \cite{Th90}.

From Eq. (\ref{13.49}) with $N_v(H)=1$, we derive the lower critical field $%
H_{c1W}$ in a finite along the layers superconductor with $W\ll \lambda _J$: 
\begin{equation}  \label{19.54}
H_{c1W}=\frac \pi {epW}=\frac{\pi ^2}2H_{c1\infty }\frac{\lambda _J}W\gg
H_{c1\infty }.
\end{equation}

For the magnetization $M=\frac 1{4\pi W}\int\limits_{-\infty }^{+\infty
}dyh(y)-\frac H{4\pi }$ we obtain: 
\begin{equation}  \label{20.55}
M(H)=-\frac{H_s^2}{16\pi H}\left[ \frac{\left| \sin \left( epWH\right)
\right| }{epWH}-(-1)^{N_v}\cos \left( epWH\right) \right] .
\end{equation}
The magnetization (\ref{20.55}) shows distinctive oscillatory behavior and
discontinuities when $epWH/\pi $ approaches an integer, i.e when a vortex
plane penetrates or leaves the superconductor. For $H\gg \Phi _0/pW$, $\frac{%
N_v\Phi _0}{pW}<$$H<\left( N_v+\frac 12\right) \frac{\Phi _0}{pW},$ the LD
superconductor exhibits a small paramagnetic effect, i.e. $M(H)>0.$ (Note
that oscillations and jumps of magnetization due to Josephson vortex
penetration have been experimentally observed on superconducting
superlattices Nb/Si \cite{Yu98}.)

\subsubsection{\bf Fraunhofer oscillations of the critical Josephson current}

Consider the case of a finite-size (along the layers) LD superconductor with 
$r(T)\ll 1$, $-L_{y1}=L_{y2}\equiv W/2$ in the presence of an externally
applied current $I$ in the $x$ direction. The boundary conditions on $\phi $
now are 
\begin{equation}  \label{56}
\frac 1{2ep}\frac{d\phi }{dy}\left( \pm \frac W2\right) =H\pm 2\pi I.
\end{equation}

Assuming $W\ll \lambda _J$, we obtain the solution up to first order in the
small parameters $r(T)$ and $W^2/\lambda _J^2$: 
\[
\phi (y)=2epHy+\pi N_v(H)+\varphi 
\]
\begin{equation}  \label{22.57}
-\frac{(-1)^{N_v}}4\frac{W^2}{\lambda _J^2}\left( epWH\right) ^{-2}\left[
\sin \left( 2epHy+\varphi \right) -2epHy\cos \left( epWH\right) \cos \varphi
-\sin \varphi \right] ,
\end{equation}
\begin{equation}  \label{23.58}
I(\varphi ,H)=\int\limits_{-\frac W2}^{+\frac W2}dyj(y)=\frac{j_0}{epH}%
\left| \sin \left( epWH\right) \right| \sin \varphi ,
\end{equation}
\begin{equation}  \label{24.59}
h(y)=H\left[ 1-\frac{(-1)^{N_v}}4\frac{W^2}{\lambda _J^2}\left( epWH\right)
^{-2}\left[ \cos \left( 2epHy+\varphi \right) -\cos \left( epWH\right) \cos
\varphi \right] \right] ,
\end{equation}
\[
f(y)=1-r(T)\left[ 1-\frac{(-1)^{N_v}\cos \left( 2epHy+\varphi \right) }{1+2%
\left[ ep\zeta (T)H\right] ^2}\right. 
\]
\[
-\frac{\sqrt{2}ep\zeta (T)H\left| \sin \left( epWH\right) \right| \cos
\varphi }{1+2\left[ ep\zeta (T)H\right] ^2}\frac{\cosh \frac{\sqrt{2}y}{%
\zeta (T)}}{\sinh \frac W{\sqrt{2}\zeta (T)}} 
\]
\begin{equation}  \label{25.60}
\left. -\frac{(-1)^{N_v}\sqrt{2}ep\zeta (T)H\cos \left( epWH\right) \sin
\varphi }{1+2\left[ ep\zeta (T)H\right] ^2}\frac{\sinh \frac{\sqrt{2}y}{%
\zeta (T)}}{\cosh \frac W{\sqrt{2}\zeta (T)}}\right] .
\end{equation}
The phase shift $\pi $$N_v(H)$, induced by $N_v$ vortex planes, assures the
condition of a minimum of the Josephson energy. The field-independent phase
shift $\varphi $ ($\left| \varphi \right| \leq \pi /2$) parameterizes the
total Josephson current $I$ given by (\ref{23.58}). Equation (\ref{23.58})
yields the well-known Fraunhofer pattern in layered superconductors \cite
{BCG92,K98}. Note that the first zero of the Fraunhofer pattern, by (\ref
{19.54}), corresponds to the lower critical field $H_{c1W}$. (See Ref. \cite
{K98} for the explanation of the Fraunhofer pattern in terms of the pinning
of the vortex planes by the edges of the superconductor.) In the absence of
the transport current, i. e., for $\varphi =0$, equations (\ref{22.57}), (%
\ref{24.59}), (\ref{25.60}) reduce, respectively, to (\ref{12.48}), (\ref
{15.50}) and (\ref{17.52}).

\subsubsection{{\bf The upper critical field $H_{c2\infty }(T)$}}

Here we consider an infinite (in the $y$ direction) LD superconductor with $%
-L_{y1}=L_{y2}\equiv W/2\rightarrow +\infty $, subject to boundary
conditions on the phase of the type (\ref{47}). Supposing that at the upper
critical field $H=H_{c2\infty }$ the transition to the normal phase is of
the second-order type, $f^2$ can be considered as a small parameter, and
equations (\ref{23}), (\ref{25}) become: 
\begin{equation}  \label{26.61}
f(y)+\zeta ^2(T)\frac{d^2f(y)}{dy^2}-r(T)\left[ 1-\cos \phi (y)\right]
f(y)=0,
\end{equation}
\begin{equation}  \label{27.62}
\frac{d^2\phi (y)}{dy^2}=0.
\end{equation}
The relevant solution of Eq. (\ref{27.62}) is 
\begin{equation}  \label{28.63}
\phi (y)=2epHy+\pi N_v(H).
\end{equation}
[Compare with (\ref{12.48}).]. The substitution of (\ref{28.63}) into (\ref
{26.61}) yields 
\begin{equation}  \label{29.64}
\frac{d^2f(t)}{dt^2}+\left[ A(T,H)-(-1)^{N_v+1}q(H)\cos 2t\right] f(t)=0,
\end{equation}
\[
A(T,H)\equiv \frac{\left[ 1-r(T)\right] }{\left[ ep\zeta (T)H\right] ^2}%
,\quad q(H)\equiv \frac{r(T)}{2\left[ ep\zeta (T)H\right] ^2}=\frac{\alpha
_{ph}}{\left[ ep\xi _{ph}H\right] ^2}, 
\]
where we have introduced a dimensionless variable $t\equiv epHy$ and the
notation $\zeta (0)\equiv \xi _{ph}$. Hence one gets two independent
equations: for the odd $N_v=2m+1$ ($m=0,1,2,\ldots $) and the even $N_v=2m$
number of vortex planes. Both of them have the usual form of the Mathieu
equations \cite{AS65}. [Note that for $N_v=2m$ Eq. (\ref{29.64}) is
well-known \cite{T96}.]

The upper critical field $H_{c2}$ is now determined by the smallest
eigenvalue of (\ref{29.64}): 
\begin{equation}  \label{31.65}
A(T,H_{c2\infty })=a_0(q_c),
\end{equation}
where $q_c\equiv q(H_{c2\infty })$, and $a_0(q)$ [$a_0(-q)=a_0(q)$] is the
smallest eigenvalue of the Mathieu equation corresponding to the
eigenfunctions $f_{N_v=2m+1}(t)\propto $ce$_0(t,q)$ and $f_{N_v=2m}(t)%
\propto $ce$_0(\pi /2-t,q)$. [Note that the function ce$_0(t,q)$ is strictly
positive and periodic with the period $\pi $.] Explicitly, equation (\ref
{31.65}) reads: 
\begin{equation}  \label{32.66}
\frac{\tau -2\alpha _{ph}}{\left[ ep\xi _{ph}H_{c2\infty }\right] ^2}%
=a_0\left( \frac{\alpha _{ph}}{\left[ ep\xi _{ph}H_{c2\infty }\right] ^2}%
\right) .
\end{equation}

Equation (\ref{32.66}) exhibits the well-known 3D-2D crossover of $%
H_{c2\infty }(T)$ \cite{T96}, with the crossover temperature determined by $%
\tau ^{*}=2\alpha _{ph}$. As usual, it is of interest to consider two
opposite limiting cases.

{\it High temperatures, weak fields}: $\tau \ll 2\alpha _{ph}$,$\quad
H_{c2\infty }\ll \sqrt{\alpha _{ph}}/ep\xi _{ph}$.

In this 3D regime, 
\begin{equation}  \label{33.67}
H_{c2\infty }(T)=\frac 1{2\sqrt{\alpha _{ph}}}\frac \tau {ep\xi _{ph}}=\frac %
1{2\sqrt{\alpha _{ph}}}\frac 1{ep\xi _{ph}}\left( 1-\frac T{T_{c0}}\right) .
\end{equation}
The superconductivity of the S-layers is strongly depressed by the vortex
planes, which can be seen by comparing local maxima $f_{\max }$ with local
minima $f_{\min }$ of the order parameter: 
\[
\frac{f_{\min }}{f_{\max }}=2\sqrt{2}\exp \left[ -2r(T)\right] \ll 1. 
\]

{\it Low temperatures, strong fields}: $2\left( 1-\frac \tau {2\alpha _{ph}}%
\right) \ll 1,\quad H_{c2\infty }\gg \sqrt{\alpha _{ph}}/ep\xi _{ph}$.

In this regime, 
\begin{equation}  \label{34.68}
H_{c2\infty }(T)=\frac{\sqrt{\alpha _{ph}}}{2ep\xi _{ph}}\left( 1-\frac \tau
{2\alpha _{ph}}\right) ^{-\frac 12}=\frac{\alpha _{ph}}{\sqrt{2}ep\xi _{ph}}%
\frac{\sqrt{T_{c0}}}{\sqrt{T-T_{c0}(1-2\alpha _{ph})}}.
\end{equation}
This expression diverges for $\tau \rightarrow \tau ^{*}-0$. The origin of
this well-known unphysical divergence is the unrealistic assumption of the
LD model of a negligible S-layer thickness. [In the microscopic theory \cite
{K98}, $H_{c2\infty }(T)$ is finite at any temperatures.] The spatial
dependence of the order parameter is given by 
\[
f(y)\propto 1-\frac{(-1)^{N_v}r(T)}{4\left[ ep\zeta (T)H_{c2\infty }\right]
^2}\cos \left( 2epH_{c2\infty }y\right) . 
\]
This spatial dependence is exactly the same as in the case of intermediate
fields (\ref{18.53}).

\section{The Meissner state in the finite LD model}

Let the LD superconductor occupy the region $\left[ L_{x1}=0\leq x\leq L_{x2}%
\right] \times \left[ L_{y1}=0\leq y\leq L_{y2}\right] $. The external
magnetic field {\bf H} ($0\leq H\leq H_s$){\bf \ }is again applied along the 
$z$ axis. The homogeneity along this axis is assumed ($W_z\rightarrow
+\infty $). The Meissner state realizes under the conditions $L_{x2}\gg
\lambda $, $L_{y2}\gg \lambda _J$, thus it is sufficient to consider the
limiting case $L_{x2}\rightarrow +\infty $, $L_{y2}\rightarrow +\infty $.

This situation is described by the functional (\ref{2}) with a minor change:
the summation is now done over $n=0,1,2,\ldots $ We assume that $r(T)\ll 1$.
Boundary conditions of the type (\ref{12}) are supposed to hold at $y=0$,
and $h(x,0)=H$. The presence of an outer boundary at $x=0$ is accounted for
by the obvious boundary condition 
\begin{equation}  \label{69}
H-h(+0,y)=\frac{pf_0^2(y)}{2e\lambda ^2(T)}\left[ \frac{d\phi _0(y) }{dy}%
-2eA_y(0,y)\right] .
\end{equation}
[Compare with (\ref{11}).] The imposition of the boundary condition (\ref{69}%
) implies a restriction on variations of $A_y(x,y)$: they must now satisfy
the condition 
\begin{equation}  \label{70}
\delta A_y(0,y)=0.
\end{equation}
The influence of the boundary at $y=0$ must vanish for $y\rightarrow +\infty 
$, hence boundary conditions 
\begin{equation}  \label{71}
\Phi _{n+1,n}(+\infty )=0,\quad \frac{d\Phi _{n+1,n}}{dy}(+\infty )=0.
\end{equation}
For $x\rightarrow +\infty $, we must arrive at the solution of the infinite
LD model (\ref{8.43})-(\ref{10.45}), thus 
\begin{equation}  \label{72}
\frac{d\phi _n(y)}{dy}-2eA_y(np,y)\rightarrow 0,\quad n\rightarrow +\infty .
\end{equation}

The minimization with respect to $f_n$ leads to (\ref{31}) and (\ref{32}).
Varying with respect to $A_x$ under the condition $\delta A_x(x,0)=\delta
A_x(x,+\infty )=0$ yields the Maxwell equations (\ref{9}) in the regions $%
np<x<(n+1)p$ ($n=0,1,2,\ldots $). Taking variations with respect to $A_y$
under the condition (\ref{70}), we obtain the Maxwell equations (\ref{10})
in the regions $np<x<(n+1)p$ ($n=0,1,2,\ldots $) and boundary conditions at
the S-layers (\ref{11}) for $n=1,2,\ldots $

The general solution of the Maxwell equations, subject to the above
formulated boundary conditions, in the gauge (\ref{21}), has the form 
\[
A(0,y)=\frac{r(T)}{4e\zeta ^2(T)}\frac 1{f_0^2(y)}\int%
\limits_0^yduf_1(u)f_0(u)\sin \phi _{1,0}(u)+\frac 1{2e}\frac{d\phi _0(y)}{dy%
}, 
\]
\[
A(x,y)=\left[ 4\pi j_0\int\limits_0^yduf_{n+1}(u)f_n(u)\sin \phi
_{n+1,n}(u)+H\right] \left[ x-\left( n+1\right) p\right] +\frac 1{2e}\frac{%
d\phi _{n+1}(y)}{dy} 
\]
\begin{equation}  \label{73}
-\frac{r(T)}{4e\zeta ^2(T)}\frac 1{f_{n+1}^2(y)}\int\limits_0^yduf_{n+1}(u)%
\left[ f_n(u)\sin \phi _{n+1,n}(u)-f_{n+2}(u)\sin \phi _{n+2,n+1}(u)\right] ,
\end{equation}
\[
np<x\leq \left( n+1\right) p,\quad n=0,1,2,\ldots , 
\]
where the phase differences $\phi _{n+1,n}$ obey the solvability conditions 
\[
\frac{d\phi _{1,0}(y)}{dy}=8\pi ej_0p\int\limits_0^yduf_1(u)f_0(u)\sin \phi
_{1,0}(u)+2epH 
\]
\[
+\frac{r(T)}{2\zeta ^2(T)}\left[ \frac 1{f_1^2(y)}\int\limits_0^yduf_1(u)%
\left[ f_0(u)\sin \phi _{1,0}(u)-f_2(u)\sin \phi _{2,1}(u)\right] \right. 
\]
\[
\left. +\frac 1{f_0^2(y)}\int\limits_0^yduf_1(u)f_0(u)\sin \phi _{1,0}(u)%
\right] , 
\]
\[
\frac{d\phi _{n+1,n}(y)}{dy}=8\pi ej_0p\int\limits_0^yduf_{n+1}(u)f_n(u)\sin
\phi _{n+1,n}(u)+2epH 
\]
\[
+\frac{r(T)}{2\zeta ^2(T)}\left[ \frac 1{f_{n+1}^2(y)}\int%
\limits_0^yduf_{n+1}(u)\left[ f_n(u)\sin \phi _{n+1,n}(u)-f_{n+2}(u)\sin
\phi _{n+2,n+1}(u)\right] \right. 
\]
\begin{equation}  \label{74}
\left. -\frac 1{f_n^2(y)}\int\limits_0^yduf_n(u)\left[ f_{n-1}(u)\sin \phi
_{n,n-1}(u)-f_{n+1}(u)\sin \phi _{n+1,n}(u)\right] \right] ,\quad
n=1,2,\ldots
\end{equation}
Equations (\ref{74}) assure the continuity of the solution (\ref{73}) at $%
x=np$ ($n=0,1,2,\ldots $). [Compare with Eqs. (\ref{a1}), (\ref{a2}) of the
infinite LD model.] The obtained solution explicitly satisfies the
current-conservation law 
\begin{equation}  \label{75}
\frac p{8\pi e\lambda ^2}\sum_{n=0}^{+\infty }f_n^2(y)\left[ \frac{d\phi
_n(y)}{dy}-2eA_y(np,y)\right] +\int\limits_0^yduj(u)=0,
\end{equation}
where $j(y)\equiv 
%TCIMACRO{\underset{n\rightarrow +\infty }{\lim }}%
%BeginExpansion
\mathrel{\mathop{\lim }\limits_{n\rightarrow +\infty }}%
%EndExpansion
j_{n+1,n}(y)$ is the density of the Josephson current given by (\ref{4.39}).
[Compare with the current-conservation law (\ref{16}) of the infinite LD
model.]

Note that, in contrast to the infinite LD model, the minimization with
respect to the phases $\phi _n$ now is not possible. Indeed, a variation of
the phase at the $m$th layer, $\delta $$\phi _m$, would induce, by (\ref{75}%
), a non-vanishing variation of the vector potential at $x=0:$%
\[
\delta A_y(0,y)=\frac 1{2e}\frac{f_m^2(y)}{f_0^2(y)}\frac{d\delta \phi _m(y) 
}{dy}. 
\]
Such a variation is not allowed by the condition (\ref{70}). This
observation explains the role of the arbitrariness contained in the system
of the Maxwell equations of the infinite LD model, discussed in the previous
section. In the infinite case, the variations $\delta A_y(\pm \infty ,y)$
were arbitrary, which allowed the system to adjust boundary conditions at $%
x\rightarrow \pm \infty $, so as to minimize the free energy with respect to
the phases.

We are interested in the behavior of (\ref{73}) in the asymptotic region $%
y\rightarrow +\infty $. The second set of the conditions (\ref{71}), applied
to (\ref{74}), in first order in the small parameter $r(T)$ yields 
\[
I_{2,1}-\left( 2+\frac{p^2}{\lambda ^2}\right) I_{1,0}-\frac{p^2H}{4\pi
\lambda ^2}=0, 
\]
\begin{equation}  \label{76}
I_{n+2,n+1}-\left( 2+\frac{p^2}{\lambda ^2}\right) I_{n+1,n}+I_{n,n-1}-\frac{%
p^2H}{4\pi \lambda ^2}=0,\quad n=1,2,\ldots ,
\end{equation}
where 
\[
I_{n+1,n}\equiv j_0\int\limits_0^{+\infty }duf_{n+1}(u)f_n(u)\sin \phi
_{n+1,n}(u),\quad n=0,1,2,\ldots 
\]
is the total Josephson current between the $(n+1)$th and the $n$th layers.
The local magnetic field inside the barriers $np<x<\left( n+1\right) p$ ($%
n=0,1,2,\ldots $) in the asymptotic region $y\rightarrow +\infty $ is given
by 
\begin{equation}  \label{77}
h_{n+1}\equiv h(x,+\infty )=4\pi I_{n+1,n}+H.
\end{equation}
With the help of the quantities $h_n$, equations (\ref{76}) can be rewritten
in the form of a recursion relation 
\begin{equation}  \label{78}
h_{n+2}-\left( 2+\frac{p^2}{\lambda ^2}\right) h_{n+1}+h_n=0,\quad
n=0,1,2,\ldots ,
\end{equation}
subject to boundary conditions 
\begin{equation}  \label{79}
h_0=H,\quad h_n\rightarrow 0,\quad n\rightarrow +\infty .
\end{equation}
The solution of (\ref{78}), (\ref{79}) is straightforward: 
\begin{equation}  \label{80}
h_n=H\left[ 1+\frac{p^2}{2\lambda ^2}-\frac p\lambda \sqrt{1+ \frac{p^2}{%
4\lambda ^2}}\right] ^n.
\end{equation}

Assuming $p\ll \lambda $, we get 
\begin{equation}  \label{81}
h(x,+\infty )=H\exp \left[ -\frac{\left( n+1\right) p}{\lambda (T)}\right]
,\quad np<x<\left( n+1\right) p,\quad n=0,1,2,\ldots
\end{equation}
The intralayer currents are given by 
\begin{equation}  \label{82}
J_n(+\infty )=\frac 1{4\pi p}\left( h_n-h_{n+1}\right) =\frac H{4\pi \lambda
(T)}\exp \left[ -\frac{np}{\lambda (T)}\right] ,\quad n=0,1,2,\ldots
\end{equation}
The order parameters are 
\begin{equation}  \label{83}
f_n(+\infty )=1-2e^2\zeta ^2(T)\lambda ^2(T)H^2\exp \left[ -\frac{2np}{%
\lambda (T)}\right] ,\quad n=0,1,2,\ldots
\end{equation}
Equations (\ref{81})- (\ref{83}) describe the Meissner state in the region $%
\left[ 0\leq x<+\infty \right) \times \left( \lambda _J\ll y<+\infty \right) 
$. In the region $\left( \lambda \ll x<+\infty \right) \times \left[ 0\leq
y<+\infty \right) $, the solution is given by (\ref{8.43})-(\ref{10.45}). As
in the case of the infinite LD model, the upper bound of the existence of
these solutions is $H=H_s$. Unfortunately, in the region $\left[ 0\leq
x<\lambda \right) \times \left[ 0\leq y<\lambda _J\right) $, an analytical
solution to Eqs. (\ref{31}) and (\ref{74}) is not possible.

Equations of the type (\ref{31}), (\ref{73}) and (\ref{74}), subject to
topological boundary conditions on $\phi _{n+1.n}$, in principle, describe
Josephson vortex configurations. In contrast to the equations of the
infinite LD model, these equations do not preclude inhomogeneous in the
layering direction vortex solutions.

\section{Discussion}

We have solved the problem of exact minimization of the LD functional in
both the infinite and the finite cases. We have shown that the LD model
belongs to a class of singular field theories \cite{D64,GT86}: the Maxwell
equations of this model contain constraints [the current conservation laws (%
\ref{16}), (\ref{75})] on the phases and the vector potential at different
superconducting layers. Such constraints, resulting from gauge invariance
combined with inherent discreteness, are typical of layered superconductors.
[See Eq. (26) of the microscopic theory \cite{K98}.] Unfortunately, the
current conservation laws (\ref{16}), (\ref{75}) were completely overlooked
in previous publications on the LD model.

By taking into account the current-conservation law (\ref{16}), we have
minimized the free energy of the infinite LD model with respect to the
phases, obtaining a closed, complete, self-consistent system of mean-field
equations (\ref{22})-(\ref{28}). We show that relations of the type (\ref{15}%
), erroneously regarded as ''equations minimizing the free energy with
respect to the phases'' \cite{BLK92}, are, in reality, mere consequences of
the Maxwell equations (\ref{9})-(\ref{11}). By considering non-vanishing
first-order variations of (\ref{2}) caused by the variation (\ref{a4}) of an
inhomogeneous solution to the Maxwell equations, we prove that the infinite
LD model does not admit solutions in the form of isolated Josephson vortices.

The exact mean-field equations (\ref{22})-(\ref{28}) contain the whole
physics of the infinite LD model in parallel magnetic fields. In particular,
they reproduce such well-known limiting cases as the Meissner state \cite
{BF92}, Fraunhofer oscillations of the critical Josephson current \cite
{BCG92}, the upper critical field $H_{c2}(T)$ \cite{KLB75,T96}, and the
vortex state \cite{Th90} in the intermediate field regime. All previous
calculations of these effects were based on a physical assumption of the
homogeneity of the solution in the layering direction. We provide a rigorous
mathematical justification of this assumption by proving that the
homogeneity of the solution is one of the necessary and sufficient
conditions of a minimum of (\ref{2}). Moreover, our approach allows us to
make self-consistent refinements on these results by obtaining exact
analytical expressions for all physical quantities of interest up to leading
order in small parameters, which substantially elucidates the physics.

We have obtained an exact topological solution (\ref{7.42})-(\ref{10.45}) to
Eqs. (\ref{22})-(\ref{28}), describing a chain of Josephson vortices (a
vortex plane). This solution clearly demonstrates that, contrary to previous
suggestions \cite{B73,CCH91}, Josephson vortices of the infinite LD model
form simultaneously and coherently (one vortex per each barrier) at the
lower critical field $H_{c1\infty }$, given by (\ref{11.46}). Successive
penetration of the vortex planes at higher fields is accompanied by
oscillations and jumps of the magnetization, as described by (\ref{20.55}).
We show that the vortex-plane solutions of the infinite LD model persist up
to the upper critical field $H_{c2\infty }(T)$ in the whole temperature
range.

Our consideration of the finite LD model illuminates the role of the
boundary effect. Thus, the imposition of the boundary condition (\ref{69})
completely eliminates all unphysical degrees of freedom and makes
minimization with respect to the phases impossible. The explicit solution (%
\ref{73}), (\ref{74}) to the Maxwell equations, in contrast to the infinite
case, does not preclude the existence of localized Josephson vortex
configurations. Making use of this solution, we obtain the first, to our
mind, self-consistent description of the Meissner state of the finite LD
model [Eqs. (\ref{81})- (\ref{83}) and (\ref{8.43})- (\ref{10.45})].

All the above results stand in full agreement with our previous
consideration of layered superconductors \cite{K98} based on a completely
different, microscopic approach. Our discussion (Appendix B) of relationship
to the microscopic theory \cite{K98} clarifies a microscopic background of
the phenomenological parameters of the LD model and casts light on its
actual domain of validity.

As regards the experimental status of the problem, simultaneous Josephson
vortex penetration into all the barriers, as described in our paper, has
been recently observed on artificial low-$T_c$ superlattices Nb/Si \cite
{Yu98}. Oscillations and jumps of the magnetization, accompanying this
penetration, have also been observed \cite{Yu98}. Concerning the reported
observation of localized Josephson vortex configurations in layered high-$%
T_c $ superconductors \cite{Mo98}, the boundary effect discussed in section
III of our paper may account for this situation. Moreover, the presence of
irregularities within the layered structure (e.g., stacking faults) can
substantially modify the physical picture: such irregularities should serve
as pinning centers for isolated Josephson vortices. We hope that our exact
results will stimulate further theoretical and experimental investigation in
these directions.

\appendix 

\section{The explicit solution to the Maxwell equations of the infinite LD
model}

In the gauge $A_x=0$, the explicit solution of Eqs. (\ref{9})-(\ref{11}) on
the intervals 
\begin{equation}  \label{a5}
(n-1)p<x\leq np,\quad n=0,\pm 1,\quad n=0,\pm 1,\pm 2,\ldots ,
\end{equation}
subject to boundary conditions (\ref{12}) at $y=L_{y1}$ and $%
h(L_{y1})=H-2\pi I$, has the form 
\[
A_y(x,y)=\left[ 4\pi j_0\int\limits_{L_{y1}}^yduf_n(u)f_{n-1}(u)\sin \phi
_{n,n-1}(u)+H-2\pi I\right] \left( x-np\right) +\frac 1{2e}\frac{d\phi _n(y) 
}{dy} 
\]
\begin{equation}  \label{a1}
-\frac{r(T)}{4e\zeta ^2(T)}\frac 1{f_n^2(y)}\int\limits_{L_{y1}}^yduf_n(u)%
\left[ f_{n-1}(u)\sin \phi _{n,n-1}(u)-f_{n+1}(u)\sin \phi _{n+1,n}(u)\right]
,
\end{equation}
where the phase differences $\phi _{n,n-1}=\phi _n-\phi _{n-1}$ obey the
solvability conditions 
\[
\frac{d\phi _{n+1,n}(y)}{dy}=8\pi
ej_0p\int\limits_{L_{y1}}^yduf_{n+1}(u)f_n(u)\sin \phi _{n+1,n}(u)+2ep\left(
H-2\pi I\right) 
\]
\[
+\frac{r(T)}{2\zeta ^2(T)}\left[ \frac 1{f_{n+1}^2(y)}\int%
\limits_{L_{y1}}^yduf_{n+1}(u)\left[ f_n(u)\sin \phi
_{n+1,n}(u)-f_{n+2}(u)\sin \phi _{n+2,n+1}(u)\right] \right. 
\]
\begin{equation}  \label{a2}
\left. -\frac 1{f_n^2(y)}\int\limits_{L_{y1}}^yduf_n(u)\left[ f_{n-1}(u)\sin
\phi _{n,n-1}(u)-f_{n+1}(u)\sin \phi _{n+1,n}(u)\right] \right] .
\end{equation}
This infinite system of integrodifferential equations assures the continuity
of the solution (\ref{a1}) at $x=np$ ($n=0,\pm 1,\pm 2,\ldots $). For $f_n=1$%
, equations (\ref{a2}) reduce to an infinite set of second-order non-linear
differential equations \cite{BC91}. Unfortunately, an explicit solution for
the vector potential of the type (\ref{a1}) was not found in previous
publications.

Consider the variation of the solution (\ref{a1}) induced by a variation of
the phase at one of the S-layers, $\delta \phi _n(y)$. As there is only one
constraint on $\frac{d\phi _n}{dy}$ and the $y$ components of the vector
potential at $x=np$, namely the current-conservation law (\ref{16}), such a
variation affects the solution (\ref{a1}) only on one of the intervals (\ref
{a5}), say, $(m-1)<x\leq mp$. Making use of (\ref{17}), we rewrite the
solution (\ref{a1}) on this interval as 
\[
A_y(x,y)=\left[ 4\pi j_0\int\limits_{L_{y1}}^yduf_m(u)f_{m-1}(u)\sin \phi
_{m,m-1}(u)+H-2\pi I\right] \left( x-mp\right) 
\]
\begin{equation}  \label{a3}
+\frac 1{2e}\frac{d\phi _m(y)}{dy}+\frac 1{f_m^2(y)}\sum_{n\not =m}f_n^2(y)%
\left[ \frac 1{2e}\frac{d\phi _n(y)}{dy}-A_y(np,y)\right] .
\end{equation}
In equation (\ref{a3}), all $\frac{d\phi _n}{dy}$ should be considered as
independent. Thus, the desired variation is 
\begin{equation}  \label{a4}
\delta A_y(x,y)=\frac 1{2e}\frac{f_n^2(y)}{f_m^2(y)}\frac{d\delta \phi _n(y)%
}{dy},
\end{equation}
where $n=0,\pm 1,\pm 2,\ldots $

\section{Relationship to the microscopic theory}

The free energy functional of the microscopic theory \cite{K98}, after the
minimization with respect to {\bf A}, has the form 
\[
\Omega \left[ f,\phi ;H\right] =\frac{H_c^2(T)}{4\pi }W_xW_z\int%
\limits_{L_{y1}}^{L_{y2}}dy\left[ \frac ap\left[ -f^2(y)+\frac 12%
f^4(y)+\zeta ^2(T)\left[ \frac{df(y)}{dy}\right] ^2\right. \right. 
\]
\[
\left. +\frac{\zeta ^2(T)}{12}\left( \frac ap\right) ^2\left[ \frac{d\phi
(y) }{dy}\right] ^2f^2(y)+\frac{\alpha \zeta ^2(T)}{a\xi _0}\left[ 1-\cos
\phi (y)\right] f^2(y)\right] 
\]

\begin{equation}  \label{b1}
\left. +4e^2\zeta ^2(T)\lambda ^2(T)\left[ \frac 1{2ep}\frac{d\phi (y)}{dy}-H%
\right] ^2\right] ,
\end{equation}
where $\xi _0$ is the BCS coherence length, $\zeta (T)$ and $\lambda (T)$
are the GL coherence length and the penetration depth, respectively, $a$ is
the S-layer thickness, 
\begin{equation}  \label{b2}
\alpha =\frac{3\pi ^2}{7\zeta (3)}\int\limits_0^1dttD(t)\ll 1,
\end{equation}
with $D(t)$ being the tunneling probability of the barrier between two
successive S-layers. The rest of notation is the same as in (\ref{22}).
Expression (\ref{b1}) applies to the temperature range (\ref{1}) and the
S-layer thicknesses meeting the condition $\xi _0\ll a\ll \zeta (T),\lambda
(T)$.

Consider the LD limit of (\ref{b1}), when $a\ll p$. In this limit, the
average kinetic energy of the intralayer currents, i.e. the term
proportional to $a^3/p^3$, should be dropped. However, the microscopic
functional (\ref{b1}) does not reduce to the corresponding LD functional (%
\ref{30}) because of the presence of the first order factor $a/p$. (In the
LD model this factor is unrealistically taken to be unity.) Nevertheless, as
can be easily seen by minimizing (\ref{b1}) with respect to $f$ and $\phi $,
the microscopic mean-field equations in this limit formally coincide with
the LD equations (\ref{22})-(\ref{28}), if one identifies $r(T)$ with the
microscopic parameter $\frac{\alpha \zeta ^2(T)}{a\xi _0}$ and the LD
quantity $j_0$ with the microscopic expression for the critical Josephson
current of a single junction 
\[
j_0=\frac{7\zeta (3)\alpha }6eN(0)\xi _0\Delta ^2(T), 
\]
where $N(0)$ is the one-spin density of states at the Fermi level.

The role of the first-order factor $a/p$ becomes evident when one considers
the penetration of an external parallel magnetic field in the layering
direction. As can be shown on the basis of the microscopic equations \cite
{K98}, the exponential falloff of the magnetic field occurs on the length
scale of the effective penetration depth $\lambda _{eff}=\lambda \sqrt{\frac %
pa}$, whereas the LD model gives $\lambda _{eff}=\lambda $. [See Eq. (\ref
{81}).]

\end{document}